\documentclass[%
 12pt, aip,notitlepage,
 amsmath,amssymb,
]{revtex4-2}

\usepackage{graphicx}
\usepackage{dcolumn}
\usepackage{bm}

\usepackage[utf8]{inputenc}
\usepackage[T1]{fontenc}
\usepackage{mathptmx}
\usepackage{etoolbox}
\usepackage{xcolor}
\usepackage[colorlinks,citecolor=blue,linkcolor=blue]{hyperref}
\usepackage[font={sf,footnotesize},labelfont={bf,footnotesize},justification=raggedright]{caption}

\makeatletter
\def\@email#1#2{%
 \endgroup
 \patchcmd{\titleblock@produce}
  {\frontmatter@RRAPformat}
  {\frontmatter@RRAPformat{\produce@RRAP{*#1\href{mailto:#2}{#2}}}\frontmatter@RRAPformat}
  {}{}
}%
\makeatother

\begin{document}

\title{Magneto-elastic softening in cold-sprayed polycrystalline nickel \\
studied by resonant ultrasound spectroscopy.}
\author{M. Janovsk\'{a}}
 \author{P. Sedl\'{a}k}
 \author{M. \v{S}ev\v{c}\'{i}k}%
 \email{hseiner@it.cas.cz}
\affiliation{ 
Institute of Thermomechanics, Czech Academy of Sciences, Dolej\v{s}kova 5, 182 00 Prague 8, Czechia}%

\author{J. Cizek}
 \affiliation{Institute of Plasma Physics, Czech Academy of Sciences, U Slovanky 2525/1, 182 00 Prague 8, Czechia}%
\affiliation{Department of Materials Engineering, Faculty of Mechanical Engineering, Czech Technical University, Karlovo namesti 13, 120 00, Prague 2, Czechia}
 
\author{J. Kondas}
\author{R. Singh}
\affiliation{ 
 Impact Innovations GmbH, B\"{u}rgermeister-Steinberger-Ring 1, Rattenkirchen,
84431, Germany
}%

\author{J. \v{C}upera} 
\affiliation{Institute of Materials Science and Engineering, Brno University of
Technology, Technick\'{a} 2896/2
616 69 Brno, Czechia}

\author{H. Seiner}
\affiliation{ 
Institute of Thermomechanics, Czech Academy of Sciences, Dolej\v{s}kova 5, 182 00 Prague 8, Czechia}%

\date{\today}

\begin{abstract}
 Cold-sprayed metallic deposits are additively manufactured materials containing high levels of compressive residual stress. Here we show that the presence and intensity of this stress can be analyzed using laser-ultrasonics, provided that the sprayed material is ferromagnetic and magnetostrictive, as in the case of pure nickel. Contactless resonant ultrasound spectroscopy is used to monitor the evolution of shear modulus and internal friction parameter of two polycrystalline Ni deposits with temperature over the Curie point, which enables a direct assessment of the strength of magneto-elastic softening that is known to be strongly stress-dependent. In addition, the proposed methodology is also shown to be suitable for in-situ observation of the recrystallization process in the vicinity of the Curie point, as well as inspecting the homogeneity of the residual stress level across the thickness of the cold-sprayed deposit. Finally, a methodology for room-temperature probing of the magnetoelastic coupling is proposed and tested on the examined materials.   
 \end{abstract}

\maketitle

\section{Introduction}

Cold spray (CS) is a cost-efficient and rapid method for deposition
of materials from feedstock powders\cite{CS1,CS2,CS3,CS4}. Initially regarded as a surface
coating method, the absence of limitation on the achievable deposit
thickness and other significant advantages of CS allowed its recent
boom as an additive manufacturing method (CSAM)\cite{AM1,AM2,AM3}. In the CS process, the powder
particles are accelerated to supersonic velocities by a pressurized gas
jet, and they undergo severe plastic deformation upon their impact
onto the substrate or onto the preceding sprayed layers. The extreme-rate
visco-plastic deformation triggers several phenomena that ultimately
lead to the formation of very strong bond between the particles through
metallurgical bonding and mechanical anchoring\cite{bonding1,bonding2,bonding3,bonding4}. As a result, the final
deposits exhibit a unique intensively-deformed microstructure as well
as significant levels of residual compressive stress\cite{Luzin_2017,Saleh_2014,Suhonen_2013}. In turn, the residual
stress affects the mechanical performance of the CS
material: at low or moderate stress
levels, its fatigue resistance is enhanced \cite{Jahed_2015,Cavaliere_2015,Sample_2020}, whereas, high levels of residual
stress, may lead to delamination of the coatings or have a detrimental impact on shapes of additively-manufactured
parts. For these reasons,
the residual stress in CS deposits and its relation to the processing
parameters have recently been extensively studied, utilizing a broad
variety of experimental techniques, ranging from in-situ curvature measurements\cite{Suhonen_2013}
and hole drilling\cite{Dayani_2018}, to utilizing neutron diffraction\cite{Loke_2014}.
\begin{figure}[!h]
\includegraphics[width=0.55\textwidth]{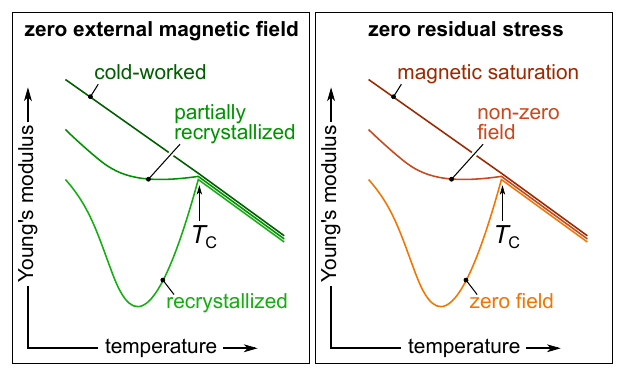}\caption{ An illustrative outline of the $\Delta E$ effect in magnetostrictive
materials. In zero magnetic field (left diagram), the level of residual
stresses is indicated by the jump in the ${\rm d}E(T)/{\rm d}T$ slope
at the Curie point $T_{\rm C}$, and the magnetoelastic softening of Young's modulus
below this temperature. If there are residual stresses suppressing
the magnetoelastic softening, the change of slope is weaker or absent.
If the initially cold-worked material undergoes recrystallization,
the level of residual stresses decreases (reaching zero for a fully
recrystallized material), which is reflected by gradual amplifying
of the $\Delta E$ effect. If there are no residual stresses (right diagram), the $\Delta E$ effect can be equivalently suppressed by performing
the $E(T)$ measurements in a magnetic field. In magnetic saturation,
the magnetization vectors do not rotate with mechanical loading, and
thus, the magnetostriction does not contribute to the strains and
does not relax part of the stresses.}
\label{fig:DeltaE}

\end{figure}

Ultrasonic methods are not typically used for such a task, despite
their emerging employment in additive manufacturing technologies\cite{ultras1,ultras2,ultras3}.
The main reason is that the elasto-acoustic constants in metals are
rather low, and thus, it is difficult to distinguish the effects of
residual stress on the acoustic parameters from those originating
from defects, such as porosity, local imperfect bonding between the
sprayed particles, or detectable volume fraction of amorphous phase
due to increased density of grain boundaries\cite{Glass_AIP_Proc_2018}. An exception from this
rule might be magnetostrictive ferromagnetic materials, in which the
residual stresses affect the elasticity through magnetoelastic coupling.
In such materials, rotations of the magnetization vector under the
action of ultrasonic vibrations lead to inelastic relaxation of a
part of the strains, which results in effective softening of the elastic
response, known as the $\Delta E$ effect. The inelastic relaxation
is suppressed by presence of residual stresses; as shown by Hubert
et al.\cite{Hubert}, as low stresses as 3 MPa can have a measurable influence on
the elastic modulus in polycrystalline nickel, and this effect becomes
even more pronounced at elevated temperatures close to the Curie point.
Figure \ref{fig:DeltaE} outlines the behavior reported and discussed
for example in Refs. {\onlinecite{Bozorth}} and   \onlinecite{Ledbetter} (see also older works
referred in the latter): in terms of its temperature dependence of the elastic modulus $E(T)$,
cold-worked polycrystalline nickel in zero magnetic field behaves
very similarly to the recrystallized polycrystalline nickel in magnetic
saturation, that is, without any change of the ${\rm d}E(T)/{\rm d}T$
slope at the Curie point. With relaxation of the residual stresses
by annealing, this behavior changes, and a clear jump in the ${\rm d}E(T)/{\rm d}T$
slope develops, resulting also in a significantly softer elastic behavior
at the room temperature (RT). This means that the
strength of the $\Delta E$ effect may serve as a reliable indicator
of the residual stresses, as well as for confirmation of their relaxation
during annealing.

In this paper, we explore the ability of resonant ultrasound spectroscopy
(RUS\cite{RUS1,RUS2,RUS_21konst}) for assessing residual stresses in CS nickel coatings, and
also the capability of this method for in-situ observation of the recrystallization
process during which the residual stresses vanish and the $\Delta E$
effect is restored. We show that this approach can also be used
to evaluate the homogeneity of a CS deposit in terms of the residual stresses, proving that they are constant over the thickness; this is another valuable information from the point of view of considering CS as an additive manufacturing technology\cite{Sinclair_Addamson_2020}. Finally, we discuss the main disadvantage of the proposed approach, which is the irreversible effect of the temperature-resolved experiments on the microstructure and the properties of the deposits. We show that these experiments can be, to some extent, replaced by room-temperature RUS measurements in magnetic field.

\section{Experiment}

\subsection{Materials and sample preparation}

\begin{figure}
\centering
\includegraphics[width=0.45\textwidth]{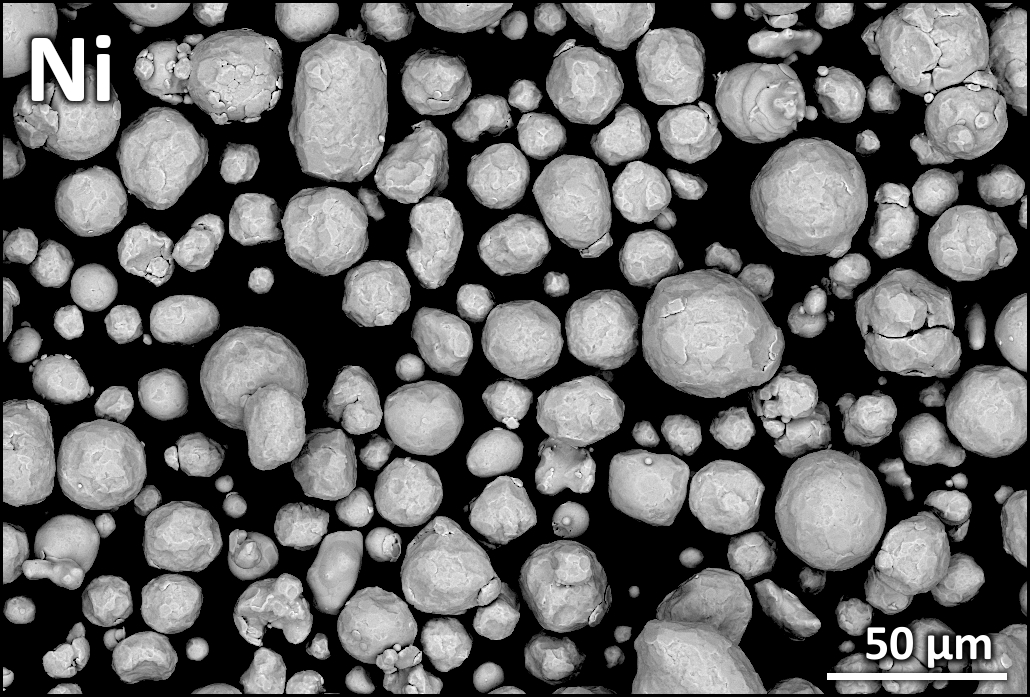}
\caption{\label{fig:Ni_powder}Morphology of the Ni powder used as a feedstock for cold spraying.}
\end{figure}

Commercial-purity Ni powder Amperit 176.068 (H.C. Starck, Goslar, Germany) was used
as the feedstock material. The powder range was measured as 15-44~$\mu$m by laser scattering (d10-d90; Mastersizer 3000, Malvern
Panalytical, UK) and its morphology was spherical (Figure \ref{fig:Ni_powder}), as a consequence
of the used gas atomization production route.
The nickel deposits were prepared using a high pressure ISS 5/11 cold
spray system (Impact Innovations, GmbH, Germany). Nitrogen was used
as the main process gas at 50 bar and 1100 $^\circ$C. The stand-off distance
was set to 30 mm and the robot arm traversal speed was 500 mm.s$^{-1}$.
Using a different number of passes, two deposits of thicknesses 6
mm and 11 mm were produced. The thinner (6 mm, denoted hereafter as
CS1) deposit was sprayed onto a 110$\times$50$\times$10 mm$^{3}$ and
the thicker deposit (11 mm, denoted hereafter as CS2) was sprayed onto a
50$\times$50$\times$10 mm$^{3}$ substrate; wrought commercial-purity
aluminum was used as the substrate material in both cases. Using the
Archimedes method, it was confirmed that both deposits had mass density
very close to bulk nickel. In particular, the density determined for
the CS1 and CS2 deposits were $\rho=(8.72\pm0.04)$ g.cm$^{-3}$ and $\rho=(8.83\pm0.04)$ g.cm$^{-3}$, respectively, giving theoretical
porosity of <1.5\%
 and <0.3\%. This also indicated slightly better compaction
in the thicker deposit, probably due to a more intenstive tamping effect triggered by the higher number 
of nozzle passes. To enable
a comparison of the magnetoelastic behavior and the $\Delta E$ for
the examined CS deposits and a well-defined bulk material, additional
experiments were done on a 4-mm sheet of low-carbon nickel (99.7\%
purity, alloy 201, ASTM B162; Bibus Metals, Czech Republic). The sheet
was hot rolled at 800 °C in the production, and had the mass density
of bulk nickel (determined as $\rho=(8.85\pm0.04)$ g.cm$^{-3}$,
a value used as the reference value for estimating the two CS porosities). 

The microstructure of all three examined materials was analyzed using
Carl Zeiss ULTRA Plus scanning electron microscope (Germany) equipped
with a Nordlys Nano EBSD detector (Oxford Instruments, England). The
analyses were performed on highly tilted samples (70$^\circ$ tilt) at an
accelerating voltage of 20 kV and an aperture of 120 $\mu$m. To estimate
the amount of plastic strains in the grains, the kernel average misorientation
(KAM) analysis was applied to the EBSD maps. 

The samples for the ultrasonic analysis were cut from the central part
of the CS1 and CS2 deposits and the reference rolled
sheet in the form of prisms with dimensions of approximately
4$\times$3$\times$2 mm$^{3}$. At least three samples were cut from each type of material,
which allowed us to confirm reproducibility of the results, as well
as to prepare series of materials with different heat treatments,
as explained below. The prismatic samples used for RUS analysis were
ground on SiC papers using lapping/polishing machine Logitech CL50 (Logitech Ltd., Glasgow, United Kingdom)
equipped with a jig to guarantee the parallelism of the opposite faces.
The largest face of each sample was then mirror polished, as needed
for the optical detection of vibrations in contactless RUS measurements.
Additionally, smaller samples with dimensions approximately 4$\times$3$\times$1.5
mm$^{3}$ were cut and similarly finished from the thicker (CS2) deposit
from four locations differing in their distance from the deposit-substrate
interface. These smaller samples were used to assess homogeneity of properties through the thickness of the deposit.

\subsection{Resonant Ultrasound Spectroscopy}

\subsubsection{Temperature-resolved RUS experiments}
A fully non-contact setup described in detail in Ref.~\onlinecite{RUS_21konst}
was used. The prismatic samples were placed into an evacuated chamber
with a low-pressure ($\sim$10 mbar) nitrogen atmosphere and a precise
temperature control of $\pm1{^\circ}$C. The vibrations of the sample
were excited by a pulsed infrared laser focused onto the bottom face
of the sample, and the time-domain response to this broadband excitation
was recorded by a scanning laser vibrometer from the upper, mirror-polished
face, and then processed using fast Fourier transform (FFT) to obtain the
resonant frequencies (frequency range 200 kHz -- 2 MHz) and the corresponding
modal shapes. 

Firstly, the RUS spectra were measured at room temperature (20 $^{\circ}$C).
At least 20 resonant frequencies were reliably
Identified for each sample. According to their modal shapes, the identified experimental
resonant modes (frequencies $f_{n}^{{\rm exp}}$) were paired with
calculated resonant modes (frequencies $f_{n}^{{\rm calc}}$) predicted
by a numerical simulation described in detail in Ref.~\onlinecite{RUS_21konst},
using elastic constants of bulk polycrystalline nickel as initial
guesses. These modes together with velocities of longitudinal ultrasonic
waves in directions perpendicular to the individual faces of the samples
were used as inputs for an inverse calculation of elastic constants.
The calculation was based on minimization of the objective function
\begin{equation}
F(c_{ij})=\sum_{n=1}^{N}\left(f_{n}^{{\rm exp}}-f_{n}^{{\rm calc}}(c_{ij})\right)^{2}
\end{equation}
 where $N$ is the number of the reliably paired modes, and $c_{ij}$
are the elastic constants. Transversal isotropy described by five
independent elastic constants was assumed first for the CS1 and CS2
materials, and Young's moduli in the spray direction and the direction
perpendicular to the spray direction were determined. Similar to the
previous studies\cite{Seiner_SCT,Cizek_pure_metals}, the deviation from isotropy
was negligible as the difference of the two moduli was below 1 \%
for all studied samples. This said, both deposits were assumed as
isotropic for further considerations. For the rolled sheet used  as the reference material, elastic isotropy was assumed
\textit{a priori}, and the measured resonant spectrum was in a reasonable
agreement with the spectrum calculated under this assumption.

\begin{figure}
 \centering
 \includegraphics[width=\textwidth]{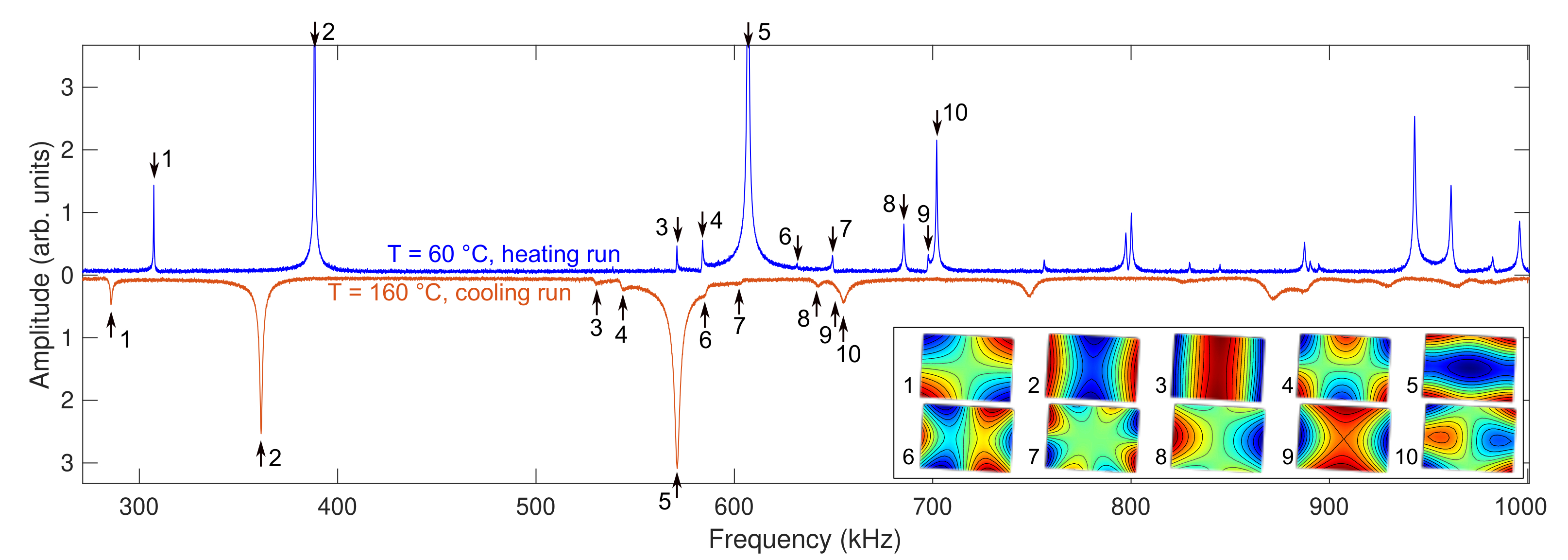}
 \caption{{}The RUS spectrum for the CS1 sample in the initial condition (upright plot) and after the thermal cycle (upside-down plot). The peaks used for tracing the $G(T)$ and $Q^{-1}(T)$ evolutions are labeled with numbers 1 to 10; the modal shapes of these modes are shown in the bottom-right corner.  }
 \label{spectrum}
\end{figure}

Secondly, to inspect the differences in the magneto-elastic
behavior of the materials, the RUS spectra for all samples were recorded in-situ during
a single temperature cycle reaching 560 $^{\circ}$C, i.e., $\sim$ 200$^{\circ}$C above the Curie temperature ($T_{{\rm C}}=354$
$^{\circ}$C for pure Ni). The measurements were performed in a chamber
equipped with a Joule-heating module, using a temperature step of
20$^{\circ}$C and an average heating rate of 2.8 $^{\circ}$C.min$^{-1}$,
including a 90s stabilization time for each measurement point. The
materials response was then recorded during
cooling with the same temperature step and the average cooling rate
of 2.5 $^{\circ}$C.min$^{-1}$. Ten dominant resonant frequencies
were traced for each sample along the temperature cycle, which enabled
accurate ($\pm1$ GPa) and reliable evaluation of the temperature
dependence of the isotropic shear modulus $G(T)$ for each material.
Changes of sample dimensions with the temperature due to thermal expansion
and the corresponding changes in density were considered, assuming
the linear expansion coefficients of pure nickel $\alpha=13\times{}10^{-6}$ K$^{-1}$.
Along the temperature cycle, the traced peaks were also used to determine
the internal friction parameter $Q^{-1}$ as  
\begin{equation}
Q^{-1}=\frac{1}{N}\sum_{n=1}^{N}\frac{{\rm FWHM}_{n}}{f_{n}},\label{eq_IF}
\end{equation}
where ${\rm FWHM}_{n}$ stands for the full width at half maximum
of the Lorentzian fit of the resonant peak at frequency $f_{n}$, and $N$ is the number of used peaks ($N=10$ was used for all materials).
It should be pointed out that the $T_{{\rm C}}$ temperature overlaps
with the temperature range in which recrystallization of strongly
cold-worked Ni can be expected\cite{Houston_1969}, and thus, the obtained $Q^{-1}(T)$ temperature dependences were expected to be a superposition of magnetoelastic damping and recrystallization-related internal friction. {{} Figure \ref{spectrum} shows examples of experimental RUS spectra used for this analysis and illustrates their evolution with temperature. It presents a comparison between a high-quality spectrum of the CS1 deposit in the initial as-sprayed condition and the spectrum of the same sample after heating it up above the recrystallization temperature and cooling back, resulting in softening of the material and a strong increase in internal friction. For the latter, the tempearature of $T=160\,^\circ$C (cooling run) was selected because it corresponded to the strongest magnetoelastic softening and highest internal friction, as discussed below. It is seen that for this temperature the peaks considerably overlapped, and thus, the ${\rm FWHM}_{n}$ for equation (\ref{eq_IF}) needed to be extracted by simultaneous fitting of the spectrum with a set of mutually overlapping Lorentzian masks. It is worth noting that the first ten resonant modes of an isotropic sample of the given geometry are dominantly determined by the shear modulus (according to the analysis reported in Ref.  \onlinecite{Nejezchlebova_NDTE_2014}, the ratio of the sensitivity of the frequencies of these modes to the shear modulus versus the bulk modulus ranges from 15:1 to 10$^4$:1), and thus, also the internal friction parameters extracted from the first ten modes can be understood as representing the shear internal friction.
This makes the $Q^{-1}$ obtained \emph{via} the relation (\ref{eq_IF}) very suitable for assessing the magnetoelastic coupling, which is a volume-preserving shear mechanism resulting from the rotation of the magnetization vector and magnetostriction.}

Finally, all materials that underwent the ${\rm RT}\rightarrow560^{\circ}{\rm C}\rightarrow{\rm RT}$
cycle were additionally annealed at 700 $^{\circ}$C for 96 hours to obtain reference samples with fully reversible and stable
$G(T)$ behaviors in the ${\rm RT}-560^{\circ}{\rm C}$ range due to their expected significant
recrystallization,
showing an amplified $\Delta E$ effect. Let us note here that although
the term $\Delta E $effect is commonly used in the literature, we
will discuss changes in the shear modulus $G$ in the following considerations,
instead of the changes of Young's modulus $E$, because the shear
modulus is more precisely determined by RUS for an isotropic material.
The changes in Young's modulus $\Delta E$ and the shear modulus $\Delta G$
are closely interconnected, and can be expressed\cite{Ledbetter}
as
\emph{
\begin{equation}
\frac{\Delta G}{G}=\left(\frac{3G}{E}\right)\frac{\Delta E}{E},\label{eq:deltaG}
\end{equation}
}
provided that the hydrostatic compressibility (the bulk modulus $K$) remains constants, which holds approximately true for magnetoelastic softening. Since 
\begin{equation}
\frac{3G}{E}=1+\frac{G}{3K},\label{eq:3GE}
\end{equation}
and since $K$ is typically much larger than
$G$ ($K=250$ GPa vs. $G=76$ GPa for bulk polycrystalline nickel
at the room temperature), the relative changes $\Delta E/E$ and $\Delta G/G$
are approximately equal to each other. 

After the ${\rm RT}\rightarrow560^{\circ}{\rm C}\rightarrow{\rm RT}$
cycle, as well as after the subsequent annealing, room-temperature RUS measurements
were repeated, to keep track of the elastic properties evolution
after the individual heat treatments.

\subsubsection{Magnetic field-resolved RUS experiments}

As an alternative approach to the temperature-resolved measurements, we further performed complementary measurements of the room-temperature (20 $^\circ$C) evolution of shear moduli of the deposits CS1 and CS2 with magnetic field. For these experiments, the contactless experimental arrangement cannot be utilized, because the magnetic field
creates a force acting on the sample, and thus, the rotation and motion
of the sample must be prevented. This can be done in medium fields
using the clamping force from ultrasonic transducers\cite{Seiner_JPCM}, or in high
magnetic field using an adhesive bonding between the transducers and
the sample\cite{Maiorov_2024}. In both cases, the individual resonant frequencies might
be shifted due to the contacts {{}or the effective inertial mass of the transducers}, but the qualitative evolution of the
elastic response with magnetic field may still enable assessing the
presence of residual stresses.

\begin{figure}
 \centering
 \includegraphics[width=0.45\textwidth]{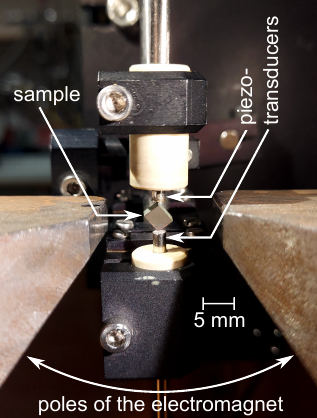}
 \caption{The experimental arrangement for field-resolved RUS measurements. The sample is clamped along its body diagonal, which prevents its rotation or motion due to increased magnetic field.}
 \label{magRUS}
\end{figure}

The experiments were performed between poles of an in-house built electromagnet, consisting of a water-cooled copper coil and a core made of ferritic steel. The electromagnet was designed for the maximum field of 0.4 T; in the used arrangement with the conventional RUS setup placed between the poles (Figure~\ref{magRUS}), the maximum reachable field was 0.26 T, which was sufficient for observing the field-induced changes of the elastic constants.

The sample in the RUS measurements was clamped between two miniature piezoelectric transducers VP-1093 (Valpey-Fisher Corp., Hopkinton, MA, USA). The excitation
was performed by a chirp signal in frequency range from 20 kHz to 1 MHz, and the output signals were detected by the acquisition system in the time domain, averaged, and transformed into the final spectra by FFT in a frequency range 200 kHz -- 1 MHz (which covers approximately the first 15 resonant frequencies for each sample). The clamping force from the transducers was enough to prevent rotations or other movements of the sample due to the magnetic field. Between the measurements at different field levels, the electromagnet was turned off to allow the coil to cool down. This ensured that the heat from the coil did not affect the temperature of the measured deposit.

In the field-resolved measurements, it was expected that the external magnetic field suppresses the $\Delta{}E$ effect similarly as the residual stress does (see the analog between the field and the stress in Figure \ref{fig:DeltaE}).  Although this falls beyond the main scope of the paper, it shows a possibility how the main disadvantages of the temperature-resolved experiments can be circumvented.  We show and discuss these results within the concluding results at the end.

\section{Results and discussion}

\subsection{Microstructure characterization}

\begin{figure}[!h]
\includegraphics[width=\textwidth]{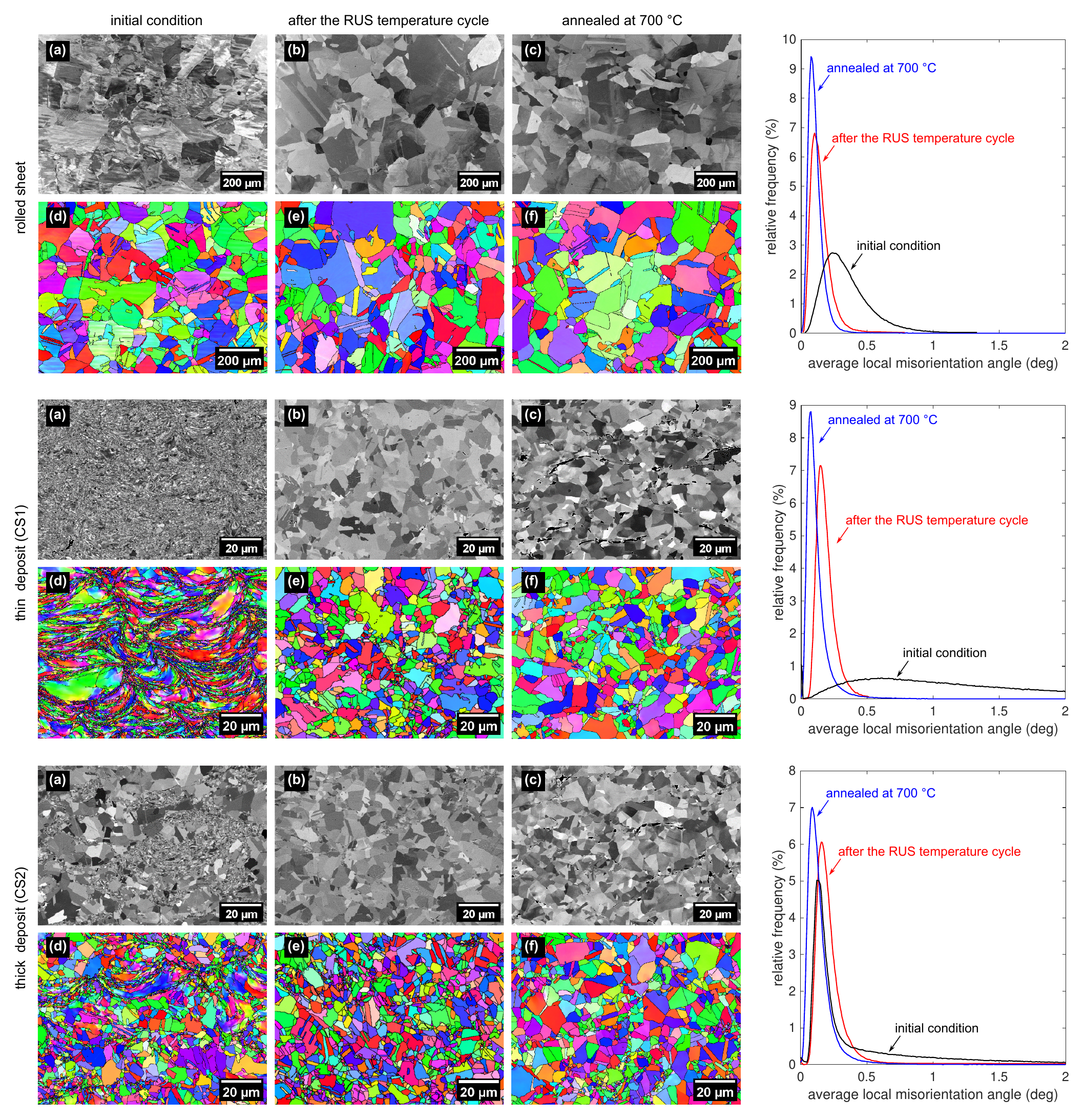}\caption{Characterization of microstructures of
the examined materials. For each material, electron channeling contrast SEM micrographs
(a,b,c) and EBSD maps (d,e,f) are shown. The micrographs and maps
refer to three conditions for each material: the initial microstructure
(a,d); the microstructure after heating the material above the Curie
point and back (b,e); and after 96~h annealing at 700~$^\circ$C (c,f).  Average misorientation angles of the lattice
inside the grains (from the KAM analysis) in the given three microstructures
for each material are shown in the right column graphs. Note the different
scalebars in the reference rolled sheet and the CS-ed deposits.}

\label{micostruc}
\end{figure}

The results of the scanning electron microscopy and the related KAM
analysis are summarized in Figure~\ref{micostruc}. For the hot-rolled
Ni sheet, large ($\text{\ensuremath{\sim} 100 \ensuremath{\mu}m}$)
grains were observed, mostly with a stripy substructure induced by the rolling, which results in relatively high misorientation angles in the KAM analysis. The
presence of this substructure indicates the presence of dislocation
walls in the rolled microstructure, and thus, residual stresses in
this material were expected. After heating the material above $T_{{\rm C}},$ the
stripy contrast disappeared, indicating recovery and/or recrystallization,
but without any visible change of the grain size. The distribution
of the KAM angles became much narrower and closer to zero, as expected
for a material recovering residual stresses. With the long-term annealing
at 700 $^{\circ}$C, this process further continued, but the resulting
changes in the SEM images as well as in the KAM curve were much less
pronounced than between the initial condition and the condition after the RUS temperature cycle. In other words, the analysis confirmed the rolled sheet
to be a suitable reference material that undergoes a transition from
plastically deformed to recovered (or recrystallized) in the two 
subsequent heat treatments. 

For the thinner CS1 deposit, the initial
condition showed a typical fine-grained CS microstructure, documenting
severe plastic deformation of the feedstock particles upon impact.
This was further confirmed by the wavy arrangement of the refined
grains as well as by the very broad distribution of the KAM angles. A bimodal grain size was observed, with chains of fine grains formed along the particle rims and relatively bigger grains retained in the particle centers, a consequence of the extreme plastic deformation during the particle impact\cite{Singh_2022}.
With the temperature cycle during the RUS measurement, the microstructure
became mostly recrystallized, with numerous micrometer-sized grains
persisted, but with the wavy pattern being fully erased by the newly
grown recrystallized grains. After annealing at 700 $^{\circ}$C,
a further grow of the recrystallized grains was observed, as confirmed
also by the KAM analysis.

For the thicker CS2 deposit, the initial microstructure
had a bimodal character, with several areas filled with ultra-fine
grains, but other consisting mainly of larger grains. This can be
interpreted as a result of partial dynamic recrystallization during
the spraying process\cite{Cs_Ni_recryst}. This deposit was sprayed onto a smaller substrate
with higher frequency of spraying passes, which resulted
in higher temperatures in the deposit during the process.
The bimodality of the microstructure in the initial condition can
be seen also from the KAM curve, which exhibits a pronounced peak
at small angles (as would be typical for a recrystallized material),
as well as a long 'tail' towards higher angles, representing the refined,
heavily plastically deformed grains. This 'tail' disappeared after
the first temperature cycle, which corresponds to the disappearance
of the smallest grains as well as of the wavy patters, clearly seen
in the SEM micrographs. Similarly as for CS1, the final annealing
in CS2 resulted in a narrower KAM angle distribution and more completely
recrystallized microstructure seen in the micrographs.

A quantitative insight into the recrystallization process in the examined
materials can be obtained from the evolution of average grain size
(measured by the average grain area in the micrograph) presented in
Table \ref{tab:Grain size}. It is seen that only for the CS1 material
the RUS temperature cycle resulted in a pronounced grain growth. This
means that only for this material, the residual stresses and the energy
of the impact-induced defects (grain refinement, substructuring)
were strong enough to trigger massive recrystallization. Further heating
for this materials stimulated further grain growth, and, as a result,
the most completely recrystallized microstructure among all studied
materials was obtained for CS1 after the final annealing at 700 $^{\circ}$C.
In contrast, nearly no grain growth was observed for the rolled sheet,
which indicated that this material did not undergo recrystallization,
and the residual stresses were relaxed by recovery of the grain substructure.
For the CS2 coating, with lower density of defects, and thus, with
lower driving force on the recrystallization, the main grain growth
was obtained during annealing at 700 $^{\circ}$C. During the RUS
temperature cycle, the smallest grains in CS2 disappeared, but the
rest of the microstructure, already partially recrystallized from
the production, remained intact, resulting in just small increase
in the grain size on average.

In summary, the microstructural
observations confirmed that the three chosen materials represent polycrystalline
nickel with three different levels of plasticity-induced grain refinement
and substructuring, and presumably also of the residual stresses,
and that these materials undergo different microstructural processes
during the used heat treatments. That is, it was confirmed that this
set of materials was suitable for analyzing the ability of RUS experiments
to monitor these features.

\begin{table*}
\caption{\label{tab:Grain size}Average grain area (in $\mu{\rm m}^{2}$) extracted
from the EBSD maps in Figure \ref{micostruc}}

\begin{tabular}{lp{1cm}cp{1cm}cp{1cm}c}
\hline {}&
 & initial condition &{}& after the RUS temperature cycle &{}& annealed at 700 °C \tabularnewline
\hline 
\hline 
rolled sheet &{} & 9.1 $\times$10$^{3}$ &{}& 10.6$\times$10$^{3}$ &{}& 11.5$\times$10$^{3}$\tabularnewline
CS1&{}  & 2.5 &{}& 9.65 &{}& 29.15\\
CS2&{}  & 5.37 &{}& 5.71 &{}& 21.32\\
\hline 
\end{tabular}
\end{table*}

\subsection{Laser-based RUS}

\subsubsection{Room-temperature elastic constants}

Room-temperature elastic constants of all studied materials and in
all discussed conditions are listed in Table~\ref{tab:RT_moduli}.
Although all materials are polycrystalline nickel with zero to very
small (<1.5\%) porosity, it is observed that these constants vary
significantly. As expected, the softest elastic behavior ($G<75$
GPa after both the RUS temperature cycle and the annealing) is observed for deposit CS1 that underwent the most
complete recrystallization, and in which the strongest $\Delta E$
effect can be expected. However, other observed differences are not
that easy to interpret. For example, in the initial condition,
the CS1 deposit is the softest among all materials, even though it
can be, based on the above microstructural analysis, expected to have
the highest level of residual stresses, and thus, the most suppressed
$\Delta E$ effect. As it will be obvious from the temperature-resolved
RUS measurements reported below, this is partially because of the
high density of defects, and this artificial softening disappears
with recrystallization; but partially, this may be also because of
the porosity. In other words, the room-temperature elasticity information
itself is insufficient for any discussion of the changes occurring
in the microstructure with the heat treatment.

\begin{table*}
\caption{\label{tab:RT_moduli}Room temperature elastic constants of nickel
sheet and CS1 and CS2 deposits after different heat treatment.}

\begin{tabular}{p{5.3cm}p{1.4cm}p{2cm}p{1.4cm}p{2cm}p{1.4cm}p{2cm}}
\hline 
 & \multicolumn{2}{l}{rolled sheet} & \multicolumn{2}{l}{CS1} & \multicolumn{2}{l}{CS2}\tabularnewline 
 & $G$ {(}GPa{)} & $E$ {(}GPa{)} & $G$ {(}GPa{)} & $E$ {(}GPa{)} & $G$ {(}GPa{)} & $E$ {(}GPa{)}\tabularnewline
\hline 
\hline 
initial condition & 80.9 & 212 & 77.7 & 203 & 79.8 & 209\tabularnewline
after the RUS temperature cycle & 79.8 & 210 & 73.8 & 195 & 78.2 & 205\tabularnewline
annealing at 700~$^{\circ}$C  & 80.5 & 212 & 74.5 & 196 & 77.0 & 202\tabularnewline
accuracy & $\pm$0.7 & $\pm$2 & $\pm$0.7 & $\pm$2 & $\pm$0.7 & $\pm$2\tabularnewline
\hline 
\end{tabular}
\end{table*}

\subsubsection{Temperature behavior}

\begin{figure*}
\includegraphics[width=\textwidth]{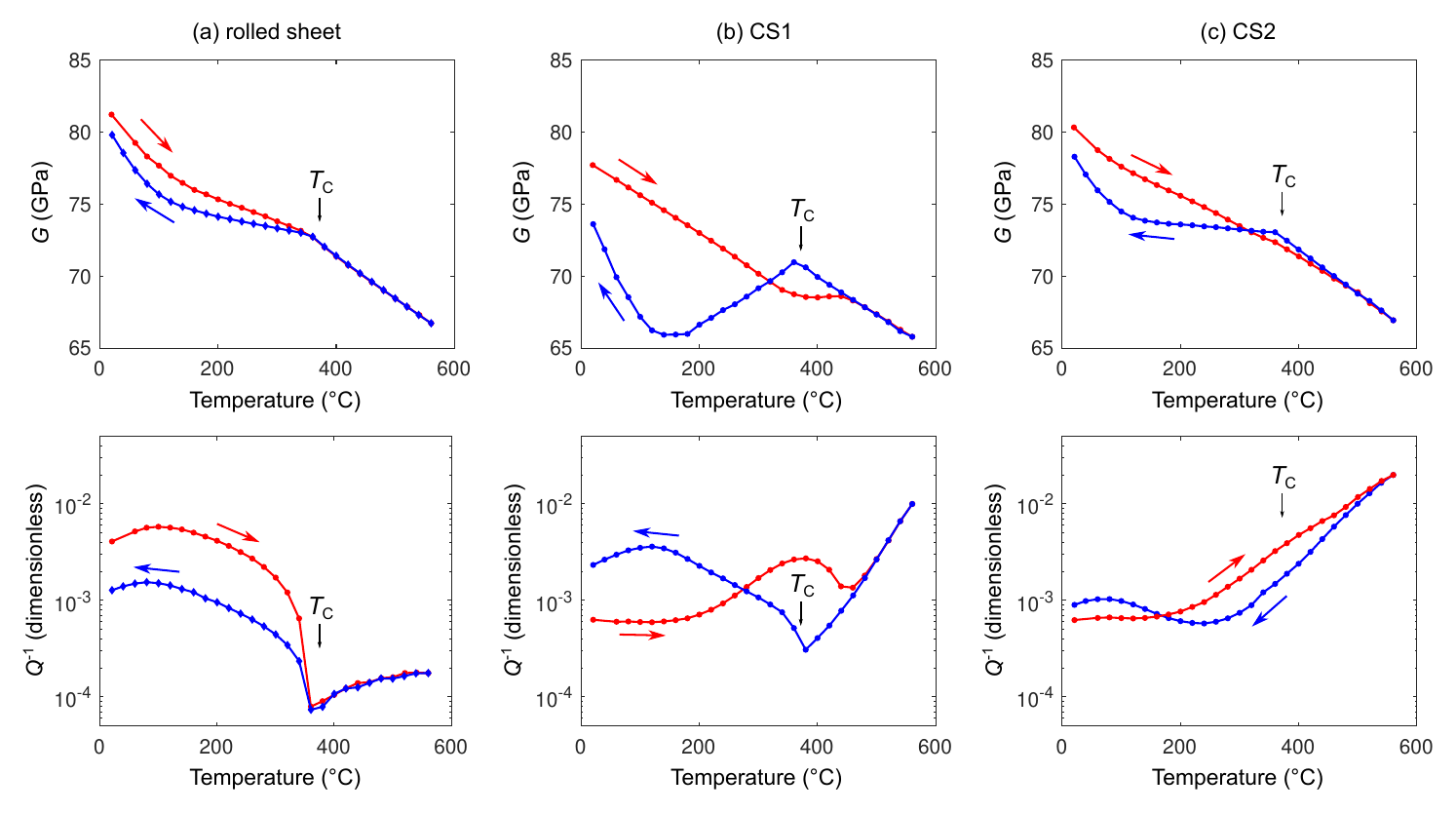}

\caption{Magnetoelastic behavior in all studied materials presented in the form
of $G(T)$ and $Q^{-1}(T)$ curves. (a) rolled Ni sheet used as a
reference bulk material; (b) thinner CS1 deposit with strong grain
refinement and wavy microstructure indicating severe plastic
deformation; (c) thicker CS2 deposit, in which partial recrystallization
occurred during the spraying process.}
\label{fig:EQ}
\end{figure*}

A more detailed insight into the magetoelastic behaviors of the
examined materials can be obtained through temperature-resolved RUS
measurements, resulting in $G(T)$ and $Q^{-1}(T)$ curves. For the
rolled sheet, these curves are shown in Figure \ref{fig:EQ}(a). The
effect of magnetism is clearly seen in both subfigures. The Curie
temperature $T_{{\rm C}}$ causes a sharp change in the ${\rm d}G(T)/{\rm d}T$
slope, as the magnetoelastic softening is active only below this temperature.
The change is sharper during the cooling run, confirming that the
relaxation of internal stresses during the temperature cycle led to
enhancing the $\Delta E$ effect. However, even this change of slope
is much less pronounced than what was reported in the literature for fully
annealed materials\cite{Hubert,Bozorth}, and it can be therefore concluded that the used
temperature cycle resulted only in a partial relaxation of the residual
stresses in the sheet. The effect of $T_{{\rm C}}$ on the internal
friction is also obvious, as the para-to-ferromagnetic transition
triggers the strong magnetoelastic damping below the Curie point.
Surprisingly, the damping is stronger in the initial condition than
after the partially release of the residual stresses. This suggests that in the initial condition, there
might be more complex coupling between the stripy substructure of
the grains and the magnetic domains, since very
fine magnetic domains may strengthen the magnetoelastic damping\cite{Seiner_JPCM}.
For both the heating and the cooling run, however, the $Q^{-1}(T)$
curves exhibit identical features. There is a maximum of the damping
at about 100 $^{\circ}$C, which probably results from an interplay
between temperature evolutions of the magnitude of magnetostriction
and of the magnetocrystalline anisotropy. And, above $T_{{\rm C}}$
the internal friction continuously increases, as expected for any
polycrystalline metal because of easier activation of nonmagnetic
relaxation processes (dislocations, point defects, grain boundary
sliding\cite{Nowick_Berry,AZ31}) with increasing temperature.

For the deposit CS1 (Figure \ref{fig:EQ}(b)), much more complex behavior
is observed for both $G(T)$ and $Q^{-1}(T)$. No sharp change of
the slope is observed for the shear modulus at $T_{{\rm C}}$ during
the heating run, which indicates the $\Delta E$ effect is completely
suppressed in the initial condition. Instead, a relative stiffening with respect to the low-temperature $G(T)$ linear trend is observed  upon further heating, terminating at approximately 480 $^{\circ}$C.
This stiffening is accompanied with a rather broad internal friction
peak. Both the stiffening and the peak indicate that recrystallization occurs in this temperature range. In the initial condition,
the severe plastic deformation resulted a high density of grain boundaries
and defect-supersaturated lattice inside the grains\cite{Cizek_2024}. The amorphous
regions inside of the boundaries as well as the heavily distorted
lattice have softer elastic constants, and thus, weak elastic stiffening
is a common feature accompanying recrystallization\cite{Benito_2005, Yu_2009, Yamaguchi_1998}. This explains
why the CS1 material is not elastically the stiffest in the initial
condition, although the $\Delta E$ effect in it is the most strongly
suppressed. The cooling curves for both $G(T)$ and $Q^{-1}(T)$ are
then very similar to those for the rolled sheet, but with a much sharper
change of the ${\rm d}G(T)/{\rm d}T$ slope at the Curie point, which
indicates a strong $\Delta E$ effect in the recrystallized microstructure.
The cooling $G(T)$ curve clearly resembles theoretically predicted
curves for polycrystalline nickel with low residual stresses, reported
in Ref. \onlinecite{Hubert}, as well as experimental curves for partially recrystallized
cold-worked materials reported in Ref.\onlinecite{Bozorth}.

For the deposit CS2 (Figure \ref{fig:EQ}(c)), the effect is qualitatively
similar as for CS1, but with several important quantitative differences.
Most importantly, there already is some weak slope change in $G(T)$ at
the Curie point during the heating run, which is in perfect agreement
with the partially recrystallized structure in the initial condition.
Then, the increase of $G$ above $T_{{\rm C}}$ due to recrystallization
is very weak (observable, but below the experimental accuracy level).
In the cooling, a stronger but still quite weak change of the slope
is observed at $T_{{\rm C}}$, which indicates that the residual stresses
were not significantly released during the cycle. On the $Q^{-1}(T)$
curve, the sharp drops at $T_{{\rm C}}$ are not observed, which
reveals that the recrystallization was triggered in this temperature
range both during the heating and the cooling runs. Because of the
minor volume fraction of the small grains and probably also a smaller
density of other defects, the recrystallization progressed much slower
than in CS1, and thus, the recrystallization peak overlapped with
the para-to-ferromagnetic transitions. 

\begin{figure}
\centering
\includegraphics[width=0.5\textwidth]{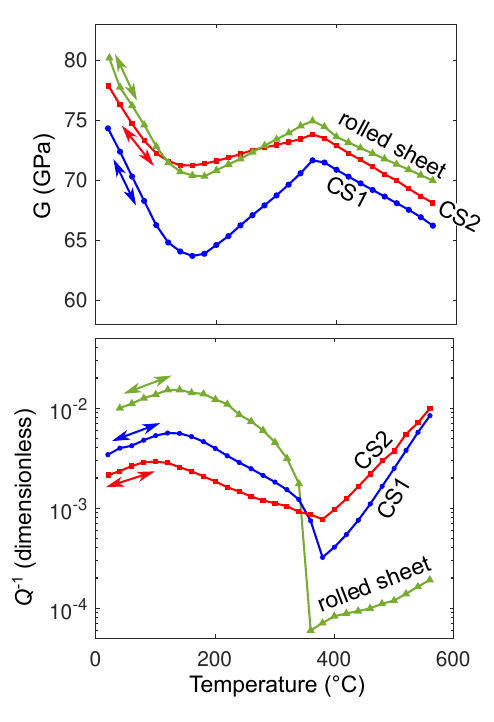}
\caption{Stable $G(T)$ and $Q^{-1}(T)$ behaviors of materials annealed at 700 $^{\circ}$C for 96~hours. }
\label{fig:annealed}
\end{figure}

To obtain stable $G(T)$ and $Q^{-1}(T)$ curves not
affected by time-dependent features related to the recrystallization for reference,
all three materials were annealed at 700 $^{\circ}$C for 96~hours.
Then, the ${\rm RT}\rightarrow560^{\circ}{\rm C}\rightarrow{\rm RT}$
RUS experiments were repeated. As expected, in this temperature range, the $G(T)$
and $Q^{-1}(T)$ became fully reversible. The behaviors are summarized
in Figure~\ref{fig:annealed}. It is seen that all $G(T)$ curves
adopted shapes typical for the $\Delta E$ effect, with the minimum
at $G(T)$ (corresponding approximately to the maximum at $Q^{-1}(T)$)
marking the temperature where the anelastic contribution to the vibrations
by magnetostriction is the strongest. Despite the annealing, however,
significant differences are still seen among the curves. The weakest
effects of $T_{{\rm C}}$ are observed for the deposit CS2, indicating
that this material retained the highest level of residual stresses.
For the rolled sheet, the effect on $G(T)$ is somehow stronger, while
the effect on $Q^{-1}(T)$ is very strong. Considering the $\sim 10^3\times$ larger
grain size and zero porosity for the sheet compared with the CS deposits,
this probably indicates that the ferromagnetic domain walls are easier
to move in the material, leading to stronger damping. This material
did not undergo recrystallization during the annealing, and thus,
the level of residual strains, as well as the $G(T)$ and $Q^{-1}(T)$
curves for it are quite similar as those measured during the cooling
run of the first ${\rm RT}\rightarrow560^{\circ}{\rm C}\rightarrow{\rm RT}$
cycle before the annealing. Finally, the strongest effect
is observed for the most completely recrystallized CS1 deposit, for
which also the elasticity at the room temperature is the softest (Table
\ref{tab:RT_moduli}). Let us point out, however, that the relative
drop in $G$ between the Curie point and its minimum, which is approximately
20\% is still much weaker than for fully recrystallized (annealed
at 1300~$^{\circ}$C) polycrystalline nickel reported in Ref.\onlinecite{Bozorth},
for which $E$ drops by more than 35\% (notice that the drops in
$E$ and $G$ should be comparable, according to Equations (\ref{eq:deltaG})
and (\ref{eq:3GE})). 

\subsubsection{Confirming homogeneity of residual stresses}

\begin{figure}
\includegraphics[width=0.5\textwidth]{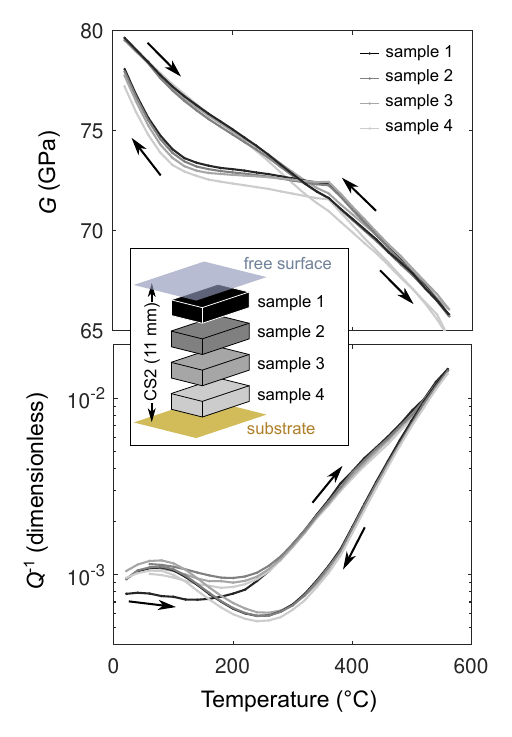}
\caption{$G(T)$ and $Q^{-1}(T)$ curves for temperature cycles reaching above
$T_{{\rm C}}$ for four samples cut from the thick deposit CS2; these
samples represent different locations across the thickness of the
deposit. The identical $G(T)$ behaviors in all these samples confirm
the homogeneity of residual stresses in the deposit. }

\label{fig:homogeneity}
\end{figure}

Since the above results prove that the magnetoelastic features on
the $G(T)$ and $Q^{-1}(T)$ curves are sensitively dependent on the
plastic strain accommodated in the microstructure, and on the related
residual stresses, the presented approach can be used to compare
different locations in the deposit. For this purpose, a set of samples
covering the entire thickness of the deposit CS2 was prepared. Each of these
samples was then subjected to the ${\rm RT}\rightarrow560^{\circ}{\rm C}\rightarrow{\rm RT}$
temperature cycle with in-situ RUS characterization. The result is
shown in Figure \ref{fig:homogeneity}. It is seen that the $G(T)$
curves for all samples are identical to each other within the experimental
accuracy, proving a perfectly constant profile of residual stresses
over the thickness of the deposit. A slight deviation in the $Q^{-1}(T)$
curve is seen for the sample from the region closest to the free surface
of the deposit (sample 1), where the internal friction is 
weaker at the very beginning of the temperature cycle. As this
deviation is not reflected in the $G(T)$ curve, it most probably
corresponds to some internal friction mechanism with much longer relaxation
times than magnetoelasticity, such as dislocation damping or grain
boundary sliding\cite{Nowick_Berry,AZ31}. These mechanisms can be somehow amplified in layers
deeper in the deposit, where grains could be more distorted due to
the tamping effect. However, the deviation is of the order of 10$^{-3}$,
and does not affect the further evolution of $Q^{-1}(T)$ with the
temperature cycle, where the curves are identical for all samples,
documenting the same progress of partial recrystallization and the
same magnetoelastic damping.

\section{Concluding remarks}

The presented results confirm that the magnetoelastic coupling and the
related $\Delta E$ effect can serve as suitable indicators of residual
stress levels in ferromagnetic cold-sprayed deposits, and that the
RUS characterization can be used as a tool for monitoring their evolution.
Using the laser-based modification of RUS, we were able to analyze
not only the shear modulus changes, but also the internal friction,
which allowed us to identify, for example, the temperature range in
which the recrystallization occurs. The $G(T)$ and $Q^{-1}(T)$ curves
clearly illustrated the differences between the CS1 and CS2 deposits,
where the latter was partially recrystallized directly from the spraying
process. The differences were seen not only in the as-sprayed behavior,
where the $\Delta E$ effect in CS1 was completely suppressed, but
also from the reversible behavior after the medium-temperature annealing,
where CS1 displayed signs of more complete recrystallization than
 CS2. The as-sprayed behavior enabled also assessing the homogeneity
of the thick deposit CS2, as the identical $G(T)$ and $Q^{-1}(T)$ curves for various samples from this deposit confirmed that there where identical microstructural processes taking place in these samples during the thermal cycle, from which one can deduce that also the initial levels of residual stresses in  samples from different distances from the substrate were identical. {{}{In principle, moreover, the obtained $G(T)$ curves may allow for quanititative determination of the levels of residual stress in the material, utilizing the theoretical model of the magnetoelastic interaction developed in Ref.\onlinecite{Hubert}. For cold-sprayed deposits, however, it must be taken into account that the variations of $G$ are also given by recrystallization-induced stiffening, and that the macroscopic elastic moduli are also affected by porosity or the interconnection between the sprayed particles. As a result, the $G(T)$ dependence can be more complicated than what predicted by the aforementioned model; the quantitative assessment of the residual stress levels would, thus, require a prior calibration by some more direct, independent method.    

Another}} disadvantage of the applied approach is obviously in the overlapping
between the temperature cycle reaching above the Curie point $T_{{\rm C}}$,
which is required for visualizing the strength of the $\Delta E$
effect, and the temperature interval for recrystallization of the
CS deposits. This could be overcome by doing the RUS experiments in
a controlled external magnetic field, owing to the equivalence between
the field and the residual stresses, as outlined in Figure~\ref{fig:DeltaE}.
 This approach might be used also for additively manufactured
CS parts of complex shapes, where field-resolved RUS measurements
would enable confirming or refuting the presence of residual stresses
in the given part, or verifying the repeatability of the manufacturing
process by comparing the $\Delta E$ effect in series of geometrically
identical CS parts.

\begin{figure}
 \centering
 \includegraphics[width=0.5\textwidth]{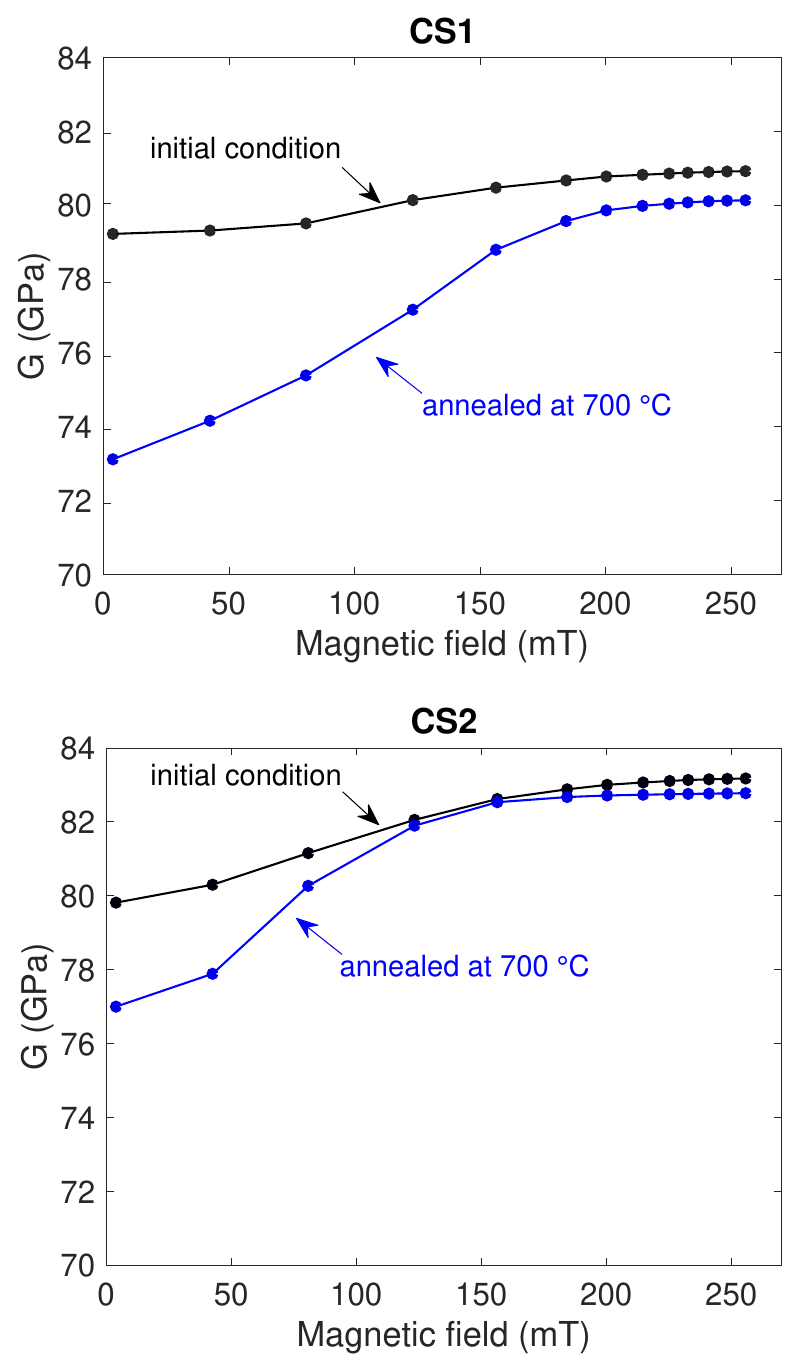}
 \caption{Room-temperature shear moduli of the deposits CS1 and CS2 measured in magnetic field, clearly illustrating the different effects of the field on materials with different microstructures and different levels of residual stresses. Notice that the zero-field values of $G$ are slightly shifted ($\lesssim$ 2 GPa) with respect to Table \ref{tab:RT_moduli}; this is because of small shifts of the resonant frequencies {{} due to moving the samples from the contact-less laser-based RUS device to the device for field-resolved experiments (Figure \ref{magRUS})}.}
 \label{magnetoelastic}
\end{figure}

As a proof of this concept, we performed field-resolved RUS experiments on samples of CS1 and CS2 deposits, each in the as-deposited condition, and after the final heat treatment. The results are shown in Figure \ref{magnetoelastic}. While the absolute values of $G(B)$ can be biased {{}by the magnetic forces} acting on the sample, {{}and possibly also by other effects resulting from the contact between the transducers and the sample\cite{Yoneda_2002,Reese_2008}}, the evolutions show exactly the expected behaviors. All materials exhibit stiffening originating from suppressing the $\Delta{E}$ effect, and the effect is the strongest for weak-to-intermediate fields, while above 0.2 T the materials are already close to magnetic saturation and $G(B)$ converges to a constant value. In both deposits, the effect of the field is stronger for the annealed materials, confirming that in the as-deposited materials, the $\Delta{E}$ effect is reduced due to residual strains.

Taking into account the {{}technical difficulties of performing the experiments in a magnetic field}, it is apparent that this type of characterization cannot bring similarly detailed insight into the processes in the studied materials and their elastic properties as obtained from temperature-resolved RUS reported above. However, the field-resolved experiments can complement their temperature-resolved counterparts, thereby enabling to qualitatively distinguish between materials with higher and lower levels of residual stresses as well.   

\subsection*{Conflict of Interest}

The authors have no conflicts of interest to disclose.

\subsection*{Data Availability}

Raw RUS spectra for all temperature resolved mesurements reported in this study are openly available in Zenodo repository at
https://doi.org/10.5281/zenodo.15743726, reference number 15743726. Other data that support the findings of this study are available from the corresponding author
upon reasonable request.

\subsection*{Acknowlegement }

This work received financial support from Czech Science Foundation
(GA\v{C}R) {[}projects Nos. 24-10334S { and 22-14048S]}, and from the Czech Ministry
of Education, Youth and Sport of the Czech Republic, within the frame
of project Ferroic Multifunctionalities (FerrMion) {[}project No.
CZ.02.01.01/00/22\textbackslash\_008/0004591{]} co-funded by the
European Union.

\end{document}